\documentstyle[pra,epsf,aps,twocolumn]{revtex}

\def\pa{{{\scriptscriptstyle p}\atop\raisebox{1ex}{$\scriptscriptstyle a$}}}
\def\pa{{{\scriptscriptstyle p}\atop\raisebox{1ex}{$\scriptscriptstyle a$}}}
\def\comment#1{}

\begin{document}
\title{Simple Explicit Formulas for Gaussian Path Integrals with Time-Dependent
Frequencies }
\author{H.\ Kleinert\thanks{%
kleinert@physik.fu-berlin.de, http://www.physik.fu-berlin.de/\~{}kleinert }
and A.\ Chervyakov\thanks{%
On leave from LCTA, JINR, Dubna, Russia}}
\address{Freie Universit\"at Berlin\\
Institut f\"ur Theoretische Physik\\
Arnimallee14, D-14195 Berlin}
\maketitle

\begin{abstract}
Quadratic fluctuations require an evaluation of ratios of functional
determinants of second-order differential operators. We relate these ratios
to the Green functions of the operators for Dirichlet, periodic and
antiperiodic boundary conditions on a line segment. This permits us to take
advantage of Wronski's construction method for Green functions without
knowledge of eigenvalues. Our final formula expresses the ratios of
functional determinants in terms of an ordinary $2\times2$ -determinant of a
constant matrix constructed from two linearly independent solutions of a the
homogeneous differential equations associated with the second-order
differential operators. For ratios of determinants encountered in
semiclassical fluctuations around a classical solution, the result can
further be expressed in terms of this classical solution.

In the presence of a zero mode, our method allows for a simple universal
regularization of the functional determinants. For Dirichlet's boundary
condition, our result is equivalent to Gelfand-Yaglom's.

Explicit formulas are given for a harmonic oscillator with an arbitrary
time-dependent frequency.
\end{abstract}

\sloppy


\section{Introduction}

Evaluation of Gaussian path integrals is needed in many physical problems,
notably in all semiclassical calculations of fluctuating systems. Typically,
we are confronted with a ratio of functional determinants of second-order
differential operators \cite{1}. For Dirichlet boundary conditions
encountered in quantum mechanical fluctuation problems, a general result has
been found by Gelfand and Yaglom \cite{2}. Working with time-sliced path
integrals, they reduced the evaluation to a simple initial-value problem for
the homogeneous second order differential equations associated with the
above operators. The functional determinants are directly given by the value
of the solutions at the final point. Unfortunately, Gelfand and Yaglom's
method becomes rather complicated for the periodic and antiperiodic boundary
conditions of quantum statistics (see Section~2.12 in \cite{1}), and has
therefore rarely been used. Several papers have studied the functional
determinants of second-order Sturm-Liouville operators with periodic
boundary conditions \cite{3}-\cite{6}, and related them to boundary-value
problems. The calculated determinants are all singular and were regularized
with the help of generalized zeta-functions \cite{7}. This has the
disadvantage of a physical quantity depending unnecessarily on the
analyticity properties of generalized zeta-functions. Moreover, the
auxiliary boundary-value problems were formulated in terms of first-order
operators, rather than the initial second-order one, making the treatment of
a zero mode of operator with periodic boundary conditions unclear, and
requiring additional work \cite{8}.

In this paper we shall avoid the above drawbacks by developing a simple and
systematic method for finding {\em ratios\/} of functional determinants of
second-order differential operators with Dirichlet, periodic and
antiperiodic boundary conditions. By focussing our attention upon ratios
instead of the determinants themselves, we avoid the need of regularization.
The main virtue of our method is that it takes advantage of the existence of
Wronski's simple construction rule for Green functions. This permits us to
reduce the functional determinants to an ordinary constant $2\times 2$
-determinant formed from solutions of homogeneous differential equations
associated with the differential operators. For semiclassical fluctuations
around a classical solution, our final result will be expressed entirely in
terms of a classical trajectory. Furthermore, for fluctuation operator with
a zero mode, a case frequently encountered in many semiclassical
calculations, we find a simple universal expression for the regularized
ratio of determinants without the zero mode.

\section{Basic Relations}

The typical fluctuation action arising in semiclassical approximations has
the form
\begin{equation}
{\cal A}[x] = \int^{t_b}_{t_a} dt\, L (\dot x(t),x(t)) = \int^{t_b}_{t_a} dt
\frac{M}{2} \left[ \dot x^2 - \Omega ^2 (t) x^2 \right].  \label{2.1}
\end{equation}
The time-dependent frequency $\Omega (t)$ can be expressed in terms of the
potential $V(x)$ of the system as
\begin{equation}
\Omega ^2 (t) = V^{\prime\prime}(x_{{\rm cl}} (t))/M,  \label{Ome@}
\end{equation}
where $x_{{\rm cl}}(t)$ is a classical trajectory solving the equation of
motion (for examples see Section~17.3 of Ref.~\cite{1}).
\begin{equation}
M \ddot x = -V^{\prime}(x).  \label{eom@}
\end{equation}
The action (\ref{2.1}) describes a harmonic oscillator with a time-dependent
frequency $\Omega (t)$. 
For this system, both the quantum mechanical propagator and the thermal
partition function contain a phase factor $\exp [i {\cal A}_{{\rm cl}}]$,
where ${\cal A}_{{\rm cl}} = {\cal A} [x_{{\rm cl}}]$ is the action of the
classical path $x_{{\rm cl}} (t)$. The phase factor is multiplied by a
fluctuation factor proportional to
\begin{equation}
F(t_b, t_a ) \sim \left( \frac{{\rm Det } K_1}{{\rm Det} \tilde K}\right) ^{
-1/2},  \label{2.2}
\end{equation}
where $K_1 = - \partial^2 _t - \Omega ^2 (t) \equiv K_0 - \Omega ^2 (t)$ is
obtained as the operator governing the second variation of the action ${\cal %
A}[x]$ along the classical path $x_{{\rm cl}} (t)$:
\begin{eqnarray}
\frac{ \delta^2 {\cal A}[x_{{\rm cl}}]}{\delta x(t)\delta x(t^{\prime})}
=\delta(t-t^{\prime})K_1.  \label{@}
\end{eqnarray}
The ratio of determinants (\ref{2.2}) arises naturally from the
normalization of the path integral \cite{1} and is well-defined. The linear
operator $K_1$ acts on the space of twice differentiable functions $y(t) =
\delta x (t)$ on an interval $t \in [t_a, t_b]$ with appropriate boundary
conditions. In the quantum-mechanical fluctuation problem, these are
Dirichlet-like with $y (t_b) = y(t_a) = 0$, in the quantum-statistical case,
they are periodic or antiperiodic with $y (t_b) = \pm y (t_a)$ and $\dot y
(t_b) = \pm \dot y (t_a)$. The operator $\tilde K$ in the denominator of (%
\ref{2.2}) may be chosen as $K_0$ (Dirichlet case) or $K_1^ \omega\equiv K_0
- \omega ^2$ (periodic and antiperiodic cases), respectively, where $\omega $
is a time-independent oscillator frequency. Then the operator $\tilde K$ is
invertible, having the Fredholm property
\begin{equation}
\frac{{\rm Det} K_1}{{\rm Det} \tilde K} = {\rm Det}\, \tilde K^{-1} K_1
\label{2.3}
\end{equation}
(a possible multiplicative anomaly being equal to unity \cite{9}).
Furthermore, since the operator $\tilde K^{-1} K_1$ is of the form $I + B$,
with $B$ an operator of the trace class, it has a well-defined determinant
without any regularization.

To calculate $F (t_b, t_a)$, we introduce a one-parameter family of
operators
\begin{equation}
K_g \equiv -\partial_{t}^2 - g \Omega ^2 (t)  \label{Kg}
\end{equation}
depending linearly on the parameter $g \in [0,1]$, and reducing to the
initial operator $K_1$ for $g=1$. Then we consider the eigen-value problem
\begin{equation}
K_g (t) y_n (g;t) = \lambda _n (g) y_n (g;t),  \label{2.4}
\end{equation}
with eigenvalues $\lambda _n (g)$. The eigenfunctions $y_n (g;t)$ satisfy
the orthonormality and completeness relations
\begin{eqnarray}
\int^{t_b}_{t_a} dt \, y_n (g;t) y_m (g;t) & = & \delta _{nm},  \label{norm@}
\\
\sum_{n} y_n (g;t) y_n (g; t^{\prime}) & = & \delta (t-t^{\prime}) .
\label{2.5}
\end{eqnarray}
The completeness relation permits us to write down immediately a spectral
representation for the Green function $G_g(t,t^{\prime})$ associated with
the differential equation (\ref{2.4}). By applying $K_g(t)$ to
\begin{equation}
G_g(t,t^{\prime})=\sum^{\infty}_{n=1}\frac{y_n(g;t)y_n(g;t^{\prime})}{
\lambda_n(g)},  \label{sprep@}
\end{equation}
and using (\ref{2.4}), (\ref{2.5}), we verify the validity of the defining
differential equation
\begin{equation}
K_g (t) G_g (t, t^{\prime}) = \delta (t-t^{\prime}).  \label{3.1}
\end{equation}

In terms of the eigenvalues $\lambda _n (g)$, the determinant (\ref{2.3})
would read
\begin{equation}
{\rm Det} \, \tilde K^{-1} K_g = C \prod_{n=1}^{\infty} \frac{ \lambda _n (g)%
}{ \lambda _n(0)} ,  \label{2.6}
\end{equation}
where $C = {\rm Det}\, (\tilde K^{-1} K_0)$ is a constant of the $g$%
-integration, which still may depend on $t_b,t_a$. Since the infinite
product of ratio of the eigenvalues $\lambda _n$ in Eq.~(\ref{2.6})
converges uniformly for all $g \in [0,1]$, we can differentiate this
equation to obtain
\begin{equation}
\partial_g \log {\rm Det} \, \tilde K^{-1} K_g = \sum_{n=1}^{\infty} \frac{
\lambda _n^{\prime}(g)}{ \lambda _n (g)}.  \label{2.7}
\end{equation}
Differentiating Eq.~(\ref{2.4}), and using the condition (\ref{2.5}) gives
for all boundary conditions,
\begin{equation}
\lambda _n^{\prime}(g) = - \int^{t_b}_{t_a} dt\, \Omega ^2 (t) y^2_n (g; t).
\label{2.8}
\end{equation}
This may be inserted into (\ref{2.7}). Because of the convergence of sum in (%
\ref{2.7}), summation and integration can be interchanged, and using the
spectral representation (\ref{sprep@}) we find the compact formula
\begin{eqnarray}
\partial_g \log {\rm Det}\, ( \tilde K^{-1} K_g) & = & - {\rm Tr}\left[
\Omega ^2 (t) G_g (t,t^{\prime})\right].  \nonumber \\
& = & - \int^{t_b}_{t_a} dt\, \Omega ^2 (t) G_g(t,t)  \label{2.10}
\end{eqnarray}
By integrating this equation in $g$, we obtain the ratio of functional
determinants (\ref{2.3}) in the form
\begin{equation}
{\rm Det}\, \tilde K^{-1} K_g = C \exp \left\{ - \int^{g}_{0}
dg^{\prime}\int^{t_b}_{t_a} dt \,\Omega ^2 (t) G_{g^{\prime}} (t,t) \right\}
\label{2.11}
\end{equation}
with the same integration constant $C$ as in Eq.~(\ref{2.6}). It is fixed
calculating the same expression for $g=0$ where the left-hand side is
well-known. In the case of Dirichlet boundary conditions where $\tilde K=K_0$%
, the left-hand side is trivially unity. For periodic and antiperiodic
boundary conditions where we take $\tilde K=K_0- \omega ^2=-\partial _t^2-
\omega ^2$, the most convenient way to normalize the right-hand side is to
go to $g=1$ and choose the frequency $\Omega^2(t)$ to be equal to the
constant frequency $\omega ^2$. The left-hand side is again unity thus
fixing $C$.

Having determined $C$ we set $g=1$ in Eq.~(\ref{2.11}) and obtain the final
result for the operator $K_1$. In the sequel we shall evaluate the
right-hand side of formula (\ref{2.11}) using explicit Wronski constructions
of the Green function for the different boundary conditions. 

\section{Wronski's Construction of Green Functions}

The general solution of the differential equation (\ref{3.1}) may be
expressed in terms of retarded and advanced Green functions which have the
general form
\begin{equation}
G_g^{(-)} (t,t^{\prime}) = G _g^{(+)} (t^{\prime}, t) = \Theta_{tt^{\prime}}
\, \Delta_g (t,t^{\prime}),  \label{3.2}
\end{equation}
where $\Theta_{tt^{\prime}} \equiv \Theta (t-t^{\prime})$ is Heaviside's
step function which vanishes for $t<t^{\prime}$ and is equal to unity for $%
t> t^{\prime}$. The function $\Delta_g (t,t^{\prime}) $ satisfies the
homogeneous differential equation corresponding to (\ref{3.1}). This is seen
by applying the operator $K_g (t)$ to (\ref{3.2}) and making use of the
identity $t \delta ^{\prime}(t) = - \delta (t)$:
\begin{eqnarray}
&&\!\!\!\!\!\!\!\!\!\!\!\!K_g (t) G_g^{(-)} (t,t^{\prime}) =
\Theta_{tt^{\prime}} \, K_g (t) \Delta_g (t,t^{\prime})  \nonumber \\
&&~~~~~ + \left[ \frac{\Delta_g (t,t^{\prime})}{(t-t^{\prime}) } - 2
\partial_t \Delta_g (t,t^{\prime}) \right] \delta (t-t^{\prime}).  \label{15}
\end{eqnarray}
Since the right-hand side must be equal to $\delta (t-t^{\prime})$, the
function $\Delta_g (t,t^{\prime}) $ has to satisfy the homogeneous
differential equation
\begin{equation}
K_g (t) \Delta_g (t,t^{\prime}) = 0,~~~{\rm for}~~ t> t^{\prime}~,
\label{3.3}
\end{equation}
while the bracket in (\ref{15}) must be equal to 1 at $t=t^{\prime}$. Upon
expanding $\Delta_g (t,t^{\prime})$ around $t=t^{\prime}$, this leads to the
conditions
\begin{equation}
\Delta_g (t,t) = 0,~~~\partial_t \Delta_g (t,t^{\prime})|_{t^{\prime}=t} =
-1 .  \label{3.4}
\end{equation}
Equation (\ref{3.3}) is solved by a linear combination
\begin{equation}
\Delta_g (t,t^{\prime}) = \alpha _g (t^{\prime}) \eta_g (t) + \beta _g
(t^{\prime}) \xi_g (t)  \label{3.5}
\end{equation}
of any two independent solutions $\eta_g (t)$ and $\xi_g (t)$ of the
homogeneous equation
\begin{equation}
K_g(t) h_g (t) = \left[ - \partial^2_t - g \Omega ^2 (t) \right] h_g (t) = 0.
\label{3.6}
\end{equation}
Their time-independent Wronski determinant $W_g = \eta_g \dot\xi_g - \dot\eta
_g \xi_g$ is nonzero, so that we can determine the coefficients in the
linear combination (\ref{3.5}) from (\ref{3.4}) and find
\begin{equation}
\Delta_g (t,t^{\prime}) = \frac{1}{W_g} \left[ \eta_g (t) \xi_g (t^{\prime})
- \xi_g (t) \eta_g (t^{\prime})\right] =- \Delta_g(t^{\prime},t).
\label{3.7}
\end{equation}
The right-hand side contains the so-called Jacobi commutator of the two
functions $\eta_g(t) $ and $\xi_g (t)$. Here we list a few algebraic
properties of $\Delta_g (t,t^{\prime})$ which will be useful in the sequel:
\begin{equation}
\Delta_g (t,t^{\prime}) = \frac{ \Delta_g (t_b,t)\Delta_g (t^{\prime}, t_a)
- \Delta_g (t_b,t^{\prime}) \Delta_g (t,t_a)}{ \Delta_g (t_a, t_b)},
\label{3.8}
\end{equation}
\begin{equation}
\Delta_g (t_b,t) \partial_{t_b} \Delta_g (t_b, t_a) - \Delta_g (t, t_a) =
\Delta_g (t_b, t_a) \partial_t \Delta_g (t_b,t),  \label{3.9}
\end{equation}
\begin{equation}
\Delta_g (t,t_a) \partial_{t_b} \Delta_g (t_b , t_a) + \Delta_g (t_b, t) =
\Delta_g (t_b, t_a) \partial_t \Delta_g (t,t_a).  \label{3.10}
\end{equation}

Note that the solution (\ref{3.2}) is so far not unique, leaving room for an
additional general solution of the homogeneous equation (\ref{3.6})
\begin{equation}
G_g (t,t^{\prime}) = \Theta _{tt^{\prime}} \Delta_g (t,t^{\prime}) + a_g
(t^{\prime}) \eta_g (t) + b_g (t^{\prime}) \xi_g (t)  \label{3.11}
\end{equation}
with arbitrary coefficients $a_g(t^{\prime})$ and $b_g(t^{\prime})$. This
ambiguity is removed by appropriate boundary conditions.

Consider first the quantum mechanical fluctuating problem with Dirichlet
boundary conditions $y (g; t_b) = y _g(t_a)=0$ for the eigenfunctions $y
(g;t)$ of $K_g$, implying for the Green function the boundary conditions
\begin{eqnarray}
G_g (t_b, t) & = & 0,~~~t_b\neq t,  \nonumber \\
G_g (t, t_a) & = & 0,~~~t \neq t_a.  \label{3.12}
\end{eqnarray}
Substituting (\ref{3.11}) into (\ref{3.12}) leads to a simple algebraic pair
of equations
\begin{eqnarray}
a_g(t) \eta_g(t_a) + b_g(t) \xi_g(t_a) & = & 0,  \label{3.13@} \\
a_g(t) \eta_g(t_b)\hspace{1pt} +\hspace{1pt} b_g(t) \xi_g(t_b) & = & -
\Delta (t_b, t).  \label{3.13}
\end{eqnarray}
We now define a fundamental matrix $\Lambda_g $ as the constant $2 \times 2$%
-matrix
\begin{equation}
\Lambda_g = \left(
\begin{array}{ll}
\eta_g(t_a) & \xi_g(t_a) \\
\eta_g(t_b) & \xi_g(t_b)
\end{array}
\right) ,  \label{3.14}
\end{equation}
and observe that under the condition
\begin{equation}
\det \Lambda _g = W_g \, \Delta_g (t_a, t_b) \neq 0,  \label{3.15}
\end{equation}
the system (\ref{3.13}) has a unique solution, so that the coefficients $%
a_g(t)$ and $b_g (t)$ in the Green function (\ref{3.11}) are easily
calculated. Making use of identity (\ref{3.8}), we obtain Wronski's
well-known formula
\begin{equation}
\! G_g (t,t^{\prime})\! =\! \frac{\Theta_{tt^{\prime}} \Delta_g (t_b,\hspace{%
-1pt} t) \Delta_g (t^{\prime}\!, \hspace{-1pt}t_a)\! + \!
\Theta_{t^{\prime}t}\Delta_g (t_b,\hspace{-1pt}t^{\prime}) \Delta_g (t,%
\hspace{-1pt}t_a) }{\Delta_g (t_a, t_b)}.  \label{3.16}
\end{equation}
For Dirichlet boundary conditions, this equation yields a unique and
well-defined Green function assuming the absence of a zero mode of the
operator $K_1$ with these boundary conditions. Such a mode would cause
problems since $\eta_1(t_a) = \eta_1(t_b )= 0$ would make $\det \Lambda _1 =
0$, thus destroying the property (\ref{3.15}) which was necessary to find (%
\ref{3.16}). Indeed, the Wronski expression (\ref{3.7}) is undetermined
since the boundary condition $\eta_1(t_a) = 0$ together with (\ref{3.13@})
imply $\xi_1(t_a) = 0$, making $W_1 = \eta_1 \dot\xi_1 - \dot\eta _1 \xi_1$
vanish at the initial time $t_a$ and thus identically in $t$.

Consider now the quantum statistical fluctuation problem with periodic or
antiperiodic boundary conditions $y (g;t_b) = \pm y(g;t_a), ~\dot y (g;t_b)
= \pm \dot y(g; t_a)$ for the eigenfunctions $y(g;t)$ of the operator $%
K_g(t) $. For the Green function $G_g^\pa (t,t^{\prime})$, these imply
\begin{eqnarray}
G_g^\pa (t_b, t^{\prime}) & = & \pm G_g^\pa (t_a, t^{\prime}),  \nonumber \\
\dot G_g^\pa (t_b,t^{\prime}) & = & \pm \dot G_g^\pa (t_a, t^{\prime}).
\label{3.17}
\end{eqnarray}
In both cases, the frequency $\Omega (t)$ and the Dirac delta function in
Eq.~(\ref{3.1}) are also assumed to be periodic or antiperiodic in time with
the same period. Inserting (\ref{3.11}) into (\ref{3.17}) gives now the
equations
\begin{eqnarray}
a(t) (\eta_b \mp \eta_a) + b(t) (\xi_b \mp \xi_a) &=& - \Delta(t_b, t),
\nonumber \\
a(t) (\dot \eta_b \mp \dot \eta_a) + b(t) (\dot \xi_b \mp \dot \xi_a)& =& -
\partial_t \Delta(t_b, t).  \label{3.18}
\end{eqnarray}
For brevity, we have omitted the subscripts $g$ and written $%
\xi_{a,b},\eta_{a,b}$ for $\xi_g(t_{a,b}), \eta_g(t_{a,b})$. Defining now
the constant $2 \times 2$ -matrices
\begin{equation}
\bar\Lambda^\pa_g = \left(
\begin{array}{ll}
\eta_b \mp \eta _a & \xi_b \mp \xi_a \\
\dot \eta_b \mp \dot\eta_a & \dot \xi_b \mp \dot \xi_a,
\end{array}
\right)  \label{3.19}
\end{equation}
the condition analogous to (\ref{3.15})
\begin{equation}
{\rm det} \,\bar \Lambda _g^\pa = W_g\, {\bar\Delta} _g^\pa(t_a,t_b) \neq 0
\label{3.20}
\end{equation}
with
\begin{equation}
{\bar\Delta} _g^\pa(t_a,t_b) = 2 \pm \partial_t \Delta_g (t_a, t_b) \pm
\partial_t \Delta_g (t_b,t_a)  \label{3.21}
\end{equation}
enables us to obtain the unique solution to Eqs.~(\ref{3.18}). After some
algebra using the identities (\ref{3.9}) and (\ref{3.10}), the expression (%
\ref{3.11}) for Green functions with periodic and antiperiodic boundary
conditions (\ref{3.17}) can be cast into the form
\begin{eqnarray}
&& G_g^\pa (t,t^{\prime}) = G_g (t,t^{\prime}) \mp  \nonumber \\
&& ~~~~~\frac{[\Delta_g (t,t_a) \pm \Delta_g (t_b,t)] [\Delta_g
(t^{\prime},t_a) \pm \Delta_g (t_b,t^{\prime})]} {{\bar\Delta}_g
^\pa(t_a,t_b) \Delta_g (t_a,t_b)}.  \label{3.22}
\end{eqnarray}
The right-hand side is well-defined unless the operator $K_1$ has a zero
mode with $\eta_b = \pm \eta_a,~ \dot \eta_b = \pm \dot \eta_a$, which would
make the determinant of the $2\times 2$ -matrix $\bar \Lambda _g^\pa$ vanish.

Note that the Green functions (\ref{3.16}) and (\ref{3.22}) are both
continuous at $t= t^{\prime}$, as is necessary for calculating the
associated ratios of functional determinants from formula (\ref{2.11}),
which we shall now do.

\section{Main Results and Relation to Gelfand-Yaglom's Initial-Value Problem}

Excluding at first zero modes, we evaluate formula (\ref{2.11}) for ratios
of functional determinants. The temporal integral on the right-hand side can
be performed efficiently following Ref.~\cite{10}. Here we present an even
more direct method, by which we express the result in terms of solutions of
Gelfand-Yaglom's initial-value problem for Dirichlet boundary conditions,
and of a dual problem for periodic and antiperiodic boundary conditions. {}~%
\newline
{}~\newline
{\bf Dirichlet Case}\newline
The Gelfand-Yaglom initial-value problem consists in the search for a
function $D_g(t)$ solving the following equations:
\begin{equation}
K_g(t) D_g(t) = 0;~~ D_g (t_a) = 0, ~~\dot D_g (t_a) = 1.  \label{4.1}
\end{equation}
By differentiating these three equations with respect to the parameter $g$,
we obtain for $D^{\prime}_g(t) \equiv \partial _g D_g(t)$ the inhomogeneous
initial-value problem
\begin{equation}
K_g(t) D^{\prime}_g(t) = \Omega ^2(t) D_g (t);~~ D^{\prime}_g (t_a) = 0,~
\dot D^{\prime}_g (t_a) = 0.  \label{4.2}
\end{equation}
The unique solution of equations (\ref{4.1}) can easily be expressed in
terms of our arbitrary set of solutions $\eta_g (t) $ and $\xi_g(t)$ as
follows
\begin{equation}
D_g(t) = \frac{\eta_g (t_a) \xi_g (t) - \xi_g (t_a) \eta_g(t)} {W_g} =
\Delta_g (t_a,t)  \label{4.3}
\end{equation}
thus leading to
\begin{equation}
D_g (t_b) = \frac{{\rm Det} \Lambda _g}{W_g} = \Delta_g(t_a,t_b).
\label{4.4}
\end{equation}
In terms of the same functions, the general solution of the inhomogeneous
initial-value problem (\ref{4.2}) can be seen to have the form
\begin{equation}
D^{\prime}_g(t) = \int^{t}_{t_a} dt^{\prime}\Omega ^2 (t^{\prime}) \Delta_g
(t,t^{\prime}) \Delta_g (t_a,t^{\prime}).  \label{4.5}
\end{equation}
Comparison with (\ref{3.16}) shows that at the final point $t=t_b$
\begin{equation}
D^{\prime}_g(t_b)=- \Delta_g(t_a,t_b) \int^{t_a}_{t_a} dt\, \Omega^2(t)
G_g(t,t).  \label{fggg}
\end{equation}
which together with (\ref{4.4}) implies the following simple relation for
the Green function (\ref{3.16}) with Dirichlet's boundary conditions:
\begin{equation}
{\rm Tr} \, [ \Omega ^2 (t) G_g (t,t^{\prime})] = - \partial_g \log \left(
\frac{{\det} \Lambda _g}{W_g}\right) = - \partial_g \log D_g (t_b).
\label{4.6}
\end{equation}
Inserting this into (\ref{2.11}), we find for the ratio of functional
determinants the simple formula
\begin{equation}
{\rm Det}\, K_0^{-1} K_g=C D_g(t_b).  \label{GY0@}
\end{equation}
The constant of integration is fixed by applying (\ref{GY0@}) to the trivial
case $g=0$, where $K_0=-\partial _t^2$ and the solution to the initial-value
problem (\ref{4.1}) is
\begin{equation}
D_0(t)=t-t_a.  \label{D0@}
\end{equation}
At $g=0$, the left-hand side of (\ref{GY@}) is unity, determining $%
C=(t_b-t_a)^{-1}$ and the final result for $g=1$:
\begin{equation}
{\rm Det}\, K_0^{-1} K_1= \frac{\det \Lambda _1}{W_1} \bigg/ \frac{{\rm Det
\Lambda _0}}{W_0}=\frac{ D_1(t_b)}{t_b-t_a}.  \label{GY@}
\end{equation}
This compact formula was first derived by Gelfand and Yaglom \cite{2} via a
direct calculation of the determinant arising in a time-sliced path
integrals \cite{1}.

{}~\newline
{\bf Periodic and Antiperiodic Case}\newline
Our technique makes it straight-forward to derive an equally compact formula
for periodic and antiperiodic boundary conditions. For this purpose we
introduce another homogeneous initial-value problem whose boundary
conditions are dual to Gelfand and Yaglom's in (\ref{4.1}):
\begin{equation}
K_g (t) \bar D_g(t) = 0;~~\bar D_g(t_a) = 1,~~ \dot{\bar D}_g(t_a) = 0.
\label{4.7}
\end{equation}
In terms of the previous arbitrary set $\eta_g(t)$ and $\xi_g(t)$ of
solutions of the homogeneous differential equation, the unique solution of (%
\ref{4.7}) reads
\begin{equation}
\bar D_g(t) = \frac{\eta _g (t) \dot\xi_g (t_a) - \xi_g (t) \dot\eta_g (t_a)
}{W_g}.  \label{4.8}
\end{equation}
This can be combined with the time derivative of (\ref{4.3}) at $t=t_b$ to
yield
\begin{equation}
\dot D_g (t_b) +\bar D_g (t_b) = \pm [2- {\bar\Delta} _g^\pa(t_a,t_b)].
\label{4.9}
\end{equation}
By differentiating Eqs.~(\ref{4.7}) with respect to $g$, we obtain the
following inhomogeneous initial-value problem for $\bar D^{\prime}_g(t) =
\partial_g \bar D_g(t)$:
\begin{equation}
K_g (t) \bar D^{\prime}_g (t) = \Omega ^2 (t) \bar D^{\prime}_g(t);~~ \bar D%
^{\prime}_g (t_a) = 0,~ \dot{\bar D^{\prime}}_g (t_a) = 0,  \label{4.10}
\end{equation}
whose general solution reads in analogy to (\ref{4.5})
\begin{equation}
\bar D^{\prime}_g (t) = - \int^{t}_{t_a} dt^{\prime}\Omega ^2 (t^{\prime})
\Delta_g (t,t^{\prime}) \dot \Delta_g (t_a,t^{\prime}),  \label{4.11}
\end{equation}
where the dot denotes the time derivative with respect of the first argument
of $\Delta_g(t,t^{\prime})$. With the help of identities (\ref{3.9}) and (%
\ref{3.10}), the combination $\dot D^{\prime}(t) + \bar D^{\prime}_g (t)$ at
$t = t_b$ can now be expressed in terms of the periodic and antiperiodic
Green functions (\ref{3.22}), in analogy to (\ref{fggg}),
\begin{equation}
\dot D^{\prime}_g (t_b) + \bar D^{\prime}_g (t_b) = \pm {\bar\Delta} _g^\pa
(t_a,t_b) \int^{t_b}_{t_a} dt \, \Omega ^2 (t) G_g^\pa (t,t).  \label{4.12}
\end{equation}
Together with (\ref{4.9}), this yields for the temporal integral on the
right-hand sides of (\ref{2.10}) and (\ref{2.11}) the simple expression
analogous to (\ref{4.6})
\begin{eqnarray}
{\rm Tr} [ \Omega ^2 (t) G_g^\pa (t,t^{\prime})] & = & - \partial_g \log
\left(\frac{ \det \bar \Lambda _g^\pa}{Wg}\right)  \nonumber \\
& = & - \partial _g\log \left[ 2 \mp \dot D_g (t_b) \mp \bar D_g (t_b)
\right].  \label{4.13}
\end{eqnarray}
This is inserted into formula Eq.~(\ref{2.11}) yielding for periodic and
antiperiodic boundary conditions
\begin{equation}
{\rm Det}\, \tilde K^{-1} K_g=C \left[ 2 \mp \dot D_g (t_b) \mp \bar D_g
(t_b) \right] ,  \label{GY01@}
\end{equation}
where $\tilde K= K_0- \omega ^2=-\partial _t^2- \omega ^2$. The constant of
integration $C$ is fixed in the way described after Eq.~(\ref{2.11}). We go
to $g=1$ and set $\Omega^2(t)\equiv \omega ^2$. For the operator $K_1^{
\omega }\equiv -\partial _t^2- \omega ^2$, we can easily solve the
Gelfand-Yaglom initial-value problem (\ref{4.1}) as well as the dual one (%
\ref{4.7}) by
\begin{equation}
D_1^{ \omega }(t)=\frac{1}{ \omega }\sin[ \omega (t-t_a)],~~~ \bar{D}_1^{
\omega }(t)=\cos[ \omega (t-t_a)],  \label{ddbar@}
\end{equation}
so that (\ref{GY01@}) determines $C$ by
\begin{equation}
\!\!\!\!1\!=\!C\left\{
\begin{array}{ll}
4\sin^2[ \omega (t_b-t_a)/2] ~~~\mbox{periodic~case,} &  \\
4\hspace{-1pt}\cos^2[ \omega (t_b-t_a)/2] ~~~\mbox{antiperiodic~case.} &
\end{array}
\right .  \label{@}
\end{equation}
Hence we obtain the final results for periodic boundary conditions
\begin{eqnarray}
{\rm Det } \, (\tilde K^{-1} K_1) & =& \frac{{\det} \bar \Lambda _1^p} {W_1} %
\bigg/ \frac{{\rm Det } \bar \Lambda ^{ \omega p}_ 1}{W_1^{\omega } }
\nonumber \\
& = & \frac{2 - \dot D_1 (t_b) - \bar D_1 (t_b) }{4\sin^2 [\omega
(t_b-t_a)/2]},  \label{4.15}
\end{eqnarray}
and for antiperiodic boundary conditions
\begin{eqnarray}
{\rm Det} \, (\tilde K^{-1} K_1)& =& \frac{{\det} \bar \Lambda _1 ^a}{W_1} %
\bigg/\frac{{\rm Det} \bar\Lambda ^{ \omega a}_ 1 }{W_1^{\omega } }
\nonumber \\
& = & \frac{2 + \dot D_1 (t_b) + \bar D_1 (t_b) }{4 \cos^2 [\omega
(t_b-t_a)/ 2]} .  \label{4.16}
\end{eqnarray}
The intermediate expressions in (\ref{GY@}), (\ref{4.15}), and (\ref{4.16})
show that the ratios of functional determinants are ordinary determinants of
two arbitrary independent solutions $\eta_1(t)$ and $\xi_1(t)$ of the
homogeneous differential equation $K_1y(t)=[-\partial_t^2 - \Omega ^2(t)]
y(t) =0$. As such, the results are manifestly invariant under arbitrary
linear transformations of these functions $(\eta_1, \xi_1) \rightarrow (%
\tilde\eta_1, \tilde \xi_1)$. 

\section{Expressions in Terms of Classical Trajectory}

In semiclassical fluctuation problems, the time-dependent frequency $%
\Omega^2(t)$ is determined by the classical solution $x_{{\rm cl}}(t)$ of
the equation of motion (\ref{eom@}) via Eq.~(\ref{Ome@}). In this case, the
above results can be made quite explicit by expressing the solutions $D_1(t)
$ and $\bar D_1(t)$ of the initial-value problems (\ref{4.1}) and (\ref{4.7}%
) directly in terms the classical trajectory $x_{{\rm cl}}(t)$ if this is
specified in terms of its initial position $x_a$ and initial velocity $\dot x%
_a$ as $x_{{\rm cl}} (t,x_a, \dot x_a)$. Given such a trajectory $x_{{\rm cl}%
} (t,x_a, \dot x_a)=x_a\bar D_1 (t)+\dot x_a D_1 (t)$ the solutions of (\ref
{4.1}) and (\ref{4.7}) can be written in the form
\begin{equation}
D_1 (t) = \frac{\partial x_{{\rm cl}} (t,x_a, \dot x_a)}{\partial \dot x_a}%
,~~ \bar D_1 (t) = \frac{\partial x_{{\rm cl}}(t,x_a, \dot x_a)}{\partial x_a%
}.  \label{4.17}
\end{equation}

As an example, take a harmonic oscillator where formulas (\ref{4.17}) are
given explicitly by the previous expressions (\ref{ddbar@}). For a classical
path, we can use the equation of motion (\ref{eom@}) and a partial
integration to express the action as a surface term
\begin{equation}
{\cal A}[x_{{\rm cl}}] = M (x_b \dot x_b - x_a \dot x_a)/2,  \label{4.17a}
\end{equation}
where
\begin{eqnarray}
x_b & = &x_a \bar D_1(t_b) + \dot x_a D_1 (t_b),  \nonumber \\
\dot x_b & =& x_a \dot{\bar D_1} (t_b) + \dot x_a \dot D_1(t_b).
\label{4.17b}
\end{eqnarray}
With the help of Eqs. (\ref{4.17b}), we can write the action (\ref{4.17a})
as a function of initial and final positions $x_a$ and $x_b$, and of the
time difference $t_b - t_a$:
\begin{eqnarray}
&&\!\!\!\!\!\!\!\!\!\!\!\!\!\!\!\!\!\!{\cal A}_{{\rm cl}} (x_a, x_b; t_b -
t_a) = \frac{M}{2D_1(t_b)}  \nonumber \\
&&\times [\dot D_1(t_b) x^2_b - 2x_b x_a + \bar D_1(t_b) x^2_a ].
\label{4.17c}
\end{eqnarray}
{}From this we obtain directly
\begin{equation}
D_1 (t_b) = -M \left[ \frac{\partial^2 {\cal A}_{{\rm cl}} (x_a, x_b, t_b -
t_a)}{\partial x_a \partial x_b} \right] ^{-1},  \label{4.18}
\end{equation}
so that the ratio (\ref{GY@}) of functional determinants for Dirichlet
boundary conditions becomes
\begin{equation}
{\rm Det}\, K_0^{-1} K_1 = -M \left[ \frac{\partial^2 {\cal A}_{{\rm cl}}
(x_a, x_b, t_b - t_a)}{\partial x_a \partial x_b} \right] ^{-1}\bigg/(t_b -
t_a).  \label{4.18a}
\end{equation}
The right-hand side is known as one-dimensional Van Vleck-Pauli-Morette
determinant (see Section~4.3 in \cite{1}).

In the case of periodic and antiperiodic boundary conditions, we find from
Eq. (\ref{4.17c})
\begin{eqnarray}
&& 2\mp\dot D_1 (t_b) \mp\bar D_1(t_b) = 2\pm \left[\frac{\partial^2 {\cal A}%
_{{\rm cl}} (x_a, x_b, t_b - t_a)}{\partial x_a \partial x_b}\right]^{-1}
\nonumber \\
&&\times\left[\frac{\partial^2 {\cal A}_{{\rm cl}} (x_a, x_b, t_b - t_a)}{%
\partial x_a^2} + \frac{\partial^2 {\cal A}_{{\rm cl}} (x_a, x_b, t_b - t_a)%
}{\partial x_b^2}\right],  \label{this}
\end{eqnarray}
which determines the ratio of functional determinants (\ref{GY01@}) in terms
of the classical action, in analogy to (\ref{4.18a}).

For a harmonic oscillator with the classical action
\begin{eqnarray}
&&\!\!\!\!\!\!\!\!\!\!\!\!{\cal A}_{{\rm cl}} (x_a, x_b, t_b - t_a)=\frac{M
\omega }{2\sin \omega (t_b-t_a)}  \nonumber \\
&&\times[(x_b^2+x_a^2)\cos \omega (t_b-t_a)-2x_bx_a],  \label{4.19}
\end{eqnarray}
and we obtain $D_1^{ \omega }(t_b)= \omega ^{-1}\sin \omega (t_b-t_a)$ as in
(\ref{ddbar@}) and
\begin{equation}
2 \mp \dot D^{\omega}_1 (t_b) \mp \bar D^{\omega}_1 (t_b) = \!\left\{
\begin{array}{ll}
\!\!4\sin^2[ \omega (t_b-t_a)/2] ~\mbox{,} &  \\
\!\!4\hspace{-1pt}\cos^2[ \omega (t_b-t_a)/2] ~\mbox{,} &
\end{array}
\right .  \label{4.21}
\end{equation}
in agreement with the previous results (\ref{GY@}), (\ref{4.15}), and (\ref
{4.16}).

\section{Treatment of Zero Mode}

Consider now the often encountered situations that the operator $K_1$ has a
zero mode. In path integrals, such a zero mode arises for example from the
translational invariance along the time axis of a classical solution in a
potential $V(x)$. As in the last section, the squared frequency $\Omega ^2
(t)$ is determined by (\ref{Ome@}).

For simplicity, we shall assume the presence of only a single zero mode,
which we choose as one of two independent solutions of the homogeneous
differential equation, say $\eta (t)$. For Dirichlet boundary conditions, we
call this a Dirichlet zero mode, satisfying
\begin{equation}
\eta _{b}=0,~~\eta _{a}=0.  \label{5.1}
\end{equation}
For periodic and antiperiodic boundary conditions, the zero mode satisfies
\begin{equation}
\eta _{b}\mp \eta _{a}=0,~~\dot{\eta}_{b}\mp \dot{\eta}_{a}=0,  \label{5.2}
\end{equation}
respectively. As pointed out earlier, the Wronski construction for
evaluating ratios of functional determinants is not applicable here since
the conditions (\ref{3.15}) and (\ref{3.20}) are violated as a consequence
of (\ref{5.1}) and (\ref{5.2}). In order to enforce (\ref{3.15}) and (\ref
{3.20}), we modify the boundary conditions for eigenfunctions $y\left(
t\right) $ of the operator $K_{1}$ by a small regulator parameter $\epsilon
>0$, and determine new eigenfunctions $y^{\epsilon }(t)$ with $y^{\epsilon
}(t)\rightarrow y(t)$ and $\lambda ^{\epsilon }\rightarrow \lambda $ for $%
\epsilon \rightarrow 0$. The specific form of regularized boundary
conditions will be irrelevant. It is merely required to keep the
boundary-value problem self-conjugated.
For instance, the Dirichlet boundary
may be slightly modified to
\begin{equation}
\eta_{a}^{\epsilon }-\epsilon \dot \eta_{a}^{\epsilon }=0,\quad
\eta_{b}^{\epsilon }+\epsilon \dot \eta_{b}^{\epsilon }=0,
\label{d1}
\end{equation}
the periodic and antiperiodic ones to
\begin{eqnarray}
\eta_{a}^{\epsilon } &=&\pm \cosh \epsilon \;\eta_{b}^{\epsilon }+\sinh
\epsilon
\,\dot\eta_{b}^{\epsilon },  \nonumber \\
\dot \eta_{a}^{\epsilon } &=&~~~\sinh \epsilon \;\eta_{b}^{\epsilon
}\pm \cosh \epsilon \stackrel{\cdot }{\eta_{b}^{\epsilon }}.  \label{d2}
\end{eqnarray}
Whereas the zero mode $\eta \left( t\right) $ satisfies
(\ref{5.1}) or (\ref{5.2}), the
modified
function
$\eta^{\epsilon }(t)$ is no longer a zero mode,
but has an eigenvalue $ \delta  \lambda ^ \epsilon $ of $K_1$,
which goes to zero for $ \epsilon \rightarrow 0$.
As long as $ \epsilon $ is nonzero, the  Wronski construction
provides us
with a regularized determinant ${\rm Det}{\,K_{1}^{\epsilon }}$
which tends to zero in the limit $\epsilon \rightarrow 0$. In terms of the
independent solutions
$\eta (t)$ and $\xi (t)$ of $K_1y=0$,
this determinant is given for the regularized
Dirichlet boundary conditions (\ref{d1}),
to first order in $\epsilon $, by
\begin{eqnarray}
&&\!{\rm Det}K_{1}^{\epsilon }=
{\rm Det}K_{1}+
\frac{\epsilon }{W}(\eta _{a}\dot{\xi}_{b}-\dot{\eta}_{a}\xi _{b}+\eta
_{b}\dot{\xi%
}_{a}-\dot{\eta}_{b}\xi _{a}).
 \label{5.3}
\end{eqnarray}
The determinant
${\rm Det}K_{1}$ vanishes, and the
constant Wronskian
\begin{equation}
W=\eta _{a}\dot{\xi}_{a}-\dot{\eta}_{a}\xi _{a}
=\eta _{b}\dot{\xi}_{b}-\dot{%
\eta}_{b}\xi _{b}  \label{5.4}
\end{equation}
 is, by (\ref{5.1}), equal to
\begin{equation}
W=-\dot{\eta}_{a}\xi _{a}=-\dot{\eta}_{b}\xi _{b}.  \label{d3}
\end{equation}
Simplifying
(\ref{5.3}) further with the help of
(\ref{5.1}), we obtain
\begin{equation}
{\rm Det}K_{1}^{\epsilon }=-\frac{\epsilon }{W}(\dot{\eta}_{a}\xi _{b}+\dot{%
\eta}_{b}\xi _{a})=\frac{\epsilon }{W^{2}}\xi _{a}\xi _{b}(\dot{\eta}%
_{b}^{2}+\dot{\eta}_{a}^{2}\;).  \label{5.5}
\end{equation}
For the regularized periodic and antiperiodic boundary conditions (\ref{d2}),
the
determinant reads,
to first order in $\epsilon $:
\begin{eqnarray}
&&{\rm Det}K_{1}^{\epsilon }=
\frac{ \epsilon }{W}(\eta
_{b}\xi _{a}-\eta _{a}\xi _{b}-\dot{\eta}_{b}\dot{\xi}_{a}+\dot{\eta}_{a}%
\dot{\xi}_{b})
,  \label{5.6}
\end{eqnarray}
with the same Wronski determinant (\ref{5.4}) whose constancy
implies,
together with (\ref{5.2}),
 that
\begin{equation}
\eta _{b}(\dot{\xi}_{b}\mp \dot{\xi}_{a})-\dot{\eta}_{b}(\xi _{b}\mp \xi
_{a})=0.  \label{5.7}
\end{equation}
Using
(\ref{5.2}) once more in (\ref{5.6}), we find
\begin{equation}
{\rm Det}K_{1}^{\epsilon }={\mp }\frac{\epsilon }{W\eta _{b}}(\eta
_{b}^{2}-\dot{\eta}_{b}^{2})(\xi _{b}\mp \xi _{a}).  \label{d4}
\end{equation}
In order to find a
finite expression for the functional determinant we must divide out
the eigenvalue $ \delta  \lambda ^ \epsilon $ before taking $ \epsilon
\rightarrow 0$.
{}From the regularized
eigenvalue equation
\begin{equation}
K_{1}\eta ^{\epsilon }(t)=\delta \lambda ^{\epsilon }\eta ^{\epsilon }(t),
\label{5.8}
\end{equation}
with $ \eta ^ \epsilon (t)$ normalized as in
(\ref{norm@}), we find to first order in $\epsilon $
\begin{eqnarray}
&&\!\!\!\!\!\!\!
\eta K_{1}\eta ^{\epsilon }=\left( \eta ^{\epsilon }\dot{\eta}-\dot{\eta}%
^{\epsilon }\eta \right) \Big|_{t_{a}}^{t_{b}}
\approx \delta \lambda ^{\epsilon
}\int_{t_{a}}^{t_{b}}dt\,\eta ^{2}(t)=\delta \lambda ^{\epsilon }.
\nonumber \label{5.9}
\end{eqnarray}
Taking into account the regularized boundary conditions (\ref{d1}) and (\ref
{d2}) for $\eta ^{\epsilon }(t)$, as well as the conditions (\ref{5.1}) and
(\ref
{5.2}) for $\eta (t)$, gives for the eigenvalue of the Dirichlet would-be
zero mode $\eta ^{\epsilon }(t)$
\begin{equation}
\delta \lambda ^{\epsilon }=\dot{\eta}_{b}\eta _{b}^{\epsilon }-\dot{\eta}%
_{a}\eta _{a}^{\epsilon }=-\epsilon (\dot{\eta}_{b}\dot{\eta}_{b}^{\epsilon
}+\dot{\eta}_{a}\dot{\eta}_{a}^{\epsilon }),  \label{5.10}
\end{equation}
and
for periodic (antiperiodic) boundary conditions:
\begin{eqnarray}
&&
\delta \lambda ^{\epsilon }=\dot{%
\eta}_{b}(\eta _{b}^{\epsilon }\mp \eta _{a}^{\epsilon })-\eta _{b}(\dot{\eta%
}_{b}^{\epsilon }\mp \dot{\eta}_{a}^{\epsilon })
=\mp  \epsilon  (\dot{\eta}_{b}\dot{\eta}_{b}^{\epsilon}
-\eta _{b}\eta _{b}^{\epsilon }).  \label{5.11}
 \end{eqnarray}
These equations enable us to remove $\delta \lambda ^{\epsilon }$ from the
regularized determinants (\ref{5.5}) and (\ref{d4}).
Defining the determinant without
the zero mode by (see Section 17.5 in \cite{1}).
\begin{equation}
{\rm Det}^{\prime }K_{1}=\lim_{\epsilon \rightarrow 0}\frac{{\rm Det}%
K_{1}^{\epsilon }}{\delta \lambda ^{\epsilon }},  \label{5.12}
\end{equation}
we obtain from
(\ref{5.5}) and (\ref{5.10})
for Dirichlet boundary conditions
\begin{eqnarray}
{{\rm Det^{\prime }}K_{1}} &=&-\frac{\xi _{b}\xi _{a}}{W^{2}}\lim_{\epsilon
\rightarrow 0}\frac{\dot{\eta}_{b}^{2}+\dot{\eta}_{a}^{2}\;}{\dot{\eta}_{b}%
\dot{\eta}_{b}^{\epsilon }+\dot{\eta}_{a}\dot{\eta}_{a}^{\epsilon }}
=-\frac{\xi _{b}\xi _{a}}{W^{2}}=-\frac{1}{\dot{\eta}_{b}\dot{\eta}_{a}}.
\label{5.13}
\end{eqnarray}
For periodic and antiperiodic boundary conditions,
the result is from (\ref{5.6}) and (\ref{5.11}):
\begin{eqnarray}
{\rm Det}^{\prime }K_{1} &=&\frac{\xi _{b}\mp \xi _{a}}{\eta _{b}W}%
\lim_{\epsilon \rightarrow 0}\frac{\eta _{b}^{2}-\dot{\eta}_{b}^{2}}{\dot{%
\eta}_{b}\dot{\eta}_{b}^{\epsilon }-\eta _{b}\eta _{b}^{\epsilon }}
=-\frac{(\xi _{b}\mp \xi _{a})}{\eta _{b}W},  \label{5.14}
\end{eqnarray}
which by (\ref{5.7}) becomes
\begin{equation}
{\rm Det}^{\prime }K_{1}=-\frac{\xi _{b}\mp \xi _{a}}{\eta _{b}W}=-\frac{%
\dot{\xi}_{b}\mp \dot{\xi}_{a}}{\dot{\eta}_{b}W}.  \label{5.15}
\end{equation}
Formulas (\ref{5.13}) and (\ref{5.15}) are useful for semiclassical
calculations of path integrals whose equations of motion
 possess
nontrivial classical solutions
such a solitons or instantons \cite{1}, as will be illustrated in Section~%
\ref{INS}.

Note that our final expressions (\ref{5.13}) and (\ref{5.15}) for the
functional
determinants are independent of the specific choice of regularization. 

\section{Time-Dependent Harmonic Oscillator}

To illustrate the power of the formulas derived in this work consider the
time-dependent harmonic oscillator described by the Lagrangian (\ref{2.1}).
The path integral formalism for such a system with the Dirichlet boundary
conditions was studied in several papers \cite{10}--\cite{12}. Here we
rederive their results and generalize them to periodic and antiperiodic
boundary conditions. Due to the absence of time-translational invariance of
the Lagrangian (\ref{2.1}), a zero mode can be excluded here. For the
Wronski construction, we take two independent solutions of Eq.~(\ref{3.6})
as follows
\begin{equation}
\eta (t) = q(t) \cos \phi (t) ,~~~ \xi (t) = q(t) \sin \phi (t)  \label{6.1}
\end{equation}
with a constant Wronski determinant $W$. The solutions $\eta (t)$ and $\xi
(t)$ are parametrized by two functions $q(t)$ and $q(t)$ satisfying the
constraint
\begin{equation}
\dot \phi(t) q^2(t) = W.  \label{6.2}
\end{equation}
The function $q(t)$ is a soliton of the Ermankov-Pinney equation \cite{13}
\begin{equation}
\ddot q + \Omega ^2 (t) q - W ^2q^{-3} = 0.  \label{6.3}
\end{equation}
For Dirichlet boundary conditions we insert (\ref{6.1}) into (\ref{GY@}),
and obtain the ratio of fluctuation determinants in the form
\begin{equation}
{\rm Det}\, K_0^{-1}K_1 =\frac{1}{W} \frac{q(t_a) q(t_b) \sin [\phi(t_b)-
\phi(t_a)]}{t_b-t_a}.  \label{6.4}
\end{equation}

For periodic or antiperiodic boundary conditions and $\Omega (t)$, the
functions $q(t)$ and $\phi (t)$ in Eq.~(\ref{6.1}) do not in general have
the same periodicity. This is possible because of the nonlinearity of Eqs. (%
\ref{6.2}) and (\ref{6.3}). Moreover, since we are assume here the absence
of a zero mode with such boundary conditions, it is a {\em necessary\/}
property of the solutions of the homogenous equations (\ref{3.6}).
Substituting (\ref{6.1}) into (\ref{4.15}) and (\ref{4.16}), we obtain the
ratios of functional determinants for periodic boundary conditions
\begin{eqnarray}
&&\!\!\!\!\!\!\!\!\!{\rm Det}\tilde{K}^{-1}K_{1}=4\sin ^{-2}\frac{\omega
(t_{b}-t_{a})}{2}  \nonumber \\
&&\!\!\!\!\!\!\!\!\!\!\times \left\{ 4\sin ^{2}\frac{[\phi (t_{b})-\phi
(t_{a})]}{2}\right.  \nonumber \\
&&\left. -\frac{[\dot{q}(t_{b})q(t_{a})-\dot{q}(t_{a})q(t_{b})]}{c}\,\sin
[\phi (t_{b})-\phi (t_{a})]\right.  \nonumber \\
&&\left. -\frac{[q(t_{b})-q(t_{a})]^{2}}{q(t_{a})q(t_{b})}\,\cos [\phi
(t_{b})-\phi (t_{a})]\right\} .  \label{6.7}
\end{eqnarray}
For antiperiodic ones, we must interchange $\sin \rightarrow -\cos $.
By a linear combination of the solutions (\ref{6.1}) we can always redefine $%
\phi (t)$ such that $\phi (t_{a})=0$.

In the literature, only formula (\ref{6.4}) for the Dirichlet case appears
to be known (see \cite{10}--\cite{12}). Formulas (\ref{6.7}) for periodic
and antiperiodic boundary conditions are new, except for predecessors in a
time-sliced formulation (see Section 2.12 in \cite{1}). The present
derivation is, however, much simpler than that of the predecessor since we
have been able to take full advantage of Wronski's simple construction
method for Green functions.

\section{Fluctuation Determinant of Instanton}

\label{INS} As an application of our formulas we derive the functional
determinant of the quadratic fluctuations around an instanton which governs
the energy level splitting of a quantum mechanical point particle in a
double-well. Setting the mass equal to unity, for simplicity, we consider a
potential of the form \cite{1}
\begin{equation}
V(x)=\frac{\omega ^{2}}{8a^{2}}\left( x^{2}-a^{2}\right) ^{2}.  \label{n1}
\end{equation}
The tunneling through the central barrier is controlled by the solution of
the equation of motion at imaginary time $\tau =-it,$ which can be
integrated once to yield the energy conservation law
\begin{equation}
\frac{1}{2}\dot{x}^{2}(\tau )=V\left( x(\tau )\right) +E,  \label{n2}
\end{equation}
where $\dot{x}\left( \tau \right) \equiv dx\left( \tau \right) /d\tau $, and
$E$ is the integration constant corresponding to the particle energy in the
inverted double-well. For the splitting between ground state and first
excited state, we must study the path integral for the evolution amplitude
over a large but finite time interval $(\tau _{a},\tau _{b})$. In a
semiclassical approximation, this is dominated by periodic solutions of with
energy $E\leq 0$, whose turning points lie close to the minima of the
double-well. We consider first a single sweep across the central barrier
from a turning point at $x(\tau _{a})=x_{a}$ to $x(\tau _{b})=x_{b}$, where
the velocities vanish: $\dot{x}(\tau _{a})=\dot{x}(\tau _{b})=0$, so that
the 
energy is given by
\begin{equation}
E=-V\left( x_{b}\right) =-V\left( x_{a}\right) =-\frac{\omega ^{2}}{8a^{2}}%
\left( x_{b}^{2}-a^{2}\right) ^{2}.  \label{n4}
\end{equation}
For a single sweep this implies
\begin{equation}
x_{b}=-x_{a},\quad x_{b}\leq a.  \label{n6}
\end{equation}
For an infinite time interval $(\tau _{a},\tau _{b})$, the sweep connects
the potential minima with each other, in which case $x_{a}=-x_{b}=a$ and $%
E=0 $ 
Then Eq.~(\ref{n2}) can easily be integrated yielding the well-known kink
solution \cite{1} centered around some finite $\tau _{0}=(\tau _{a}+\tau
_{b})/2$:
\begin{equation}
x_{{\rm {cl}}}\left( \tau \right) \,=\,a\tanh \left[ \frac{\omega \left(
\tau -\tau _{0}\right) }{2}\right] .  \label{n9}
\end{equation}
With the explicit energy (\ref{n4}), the equation of motion (\ref{n2}) reads
\begin{equation}
{\dot{x}}^{2}\left( \tau \right) =\frac{\omega ^{2}}{4a^{2}}\left(
x_{b}^{2}-x^{2}\right) \left( b^{2}-x^{2}\right) ,  \label{n10}
\end{equation}
where $b^{2}\equiv 2a^{2}-x_{b}^{2}$ and $x^{2}\leq x_{b}^{2}\leq b^{2}.$
Integrating Eq.~(\ref{n10}) gives
\begin{equation}
\int\limits_{x(\tau )}^{x_{b}}\frac{dt}{\sqrt{\left( x_{b}^{2}-t^{2}\right)
\left( b^{2}-t^{2}\right) }}=-\frac{\omega }{2a}\left( \tau -\tau
_{b}\right) .  \label{n11}
\end{equation}
It is useful to introduce a normalized coordinate $y(\tau )\equiv x\left(
\tau \right) /x_{b}$ moving between $-1$ and $1$, and rewrite (\ref{n11}) as
\begin{equation}
\frac{1}{b}\int\limits_{0}^{y\left( \tau \right) }\frac{dt}{\sqrt{\left(
1-t^{2}\right) \left( 1-m\,t^{2}\right) }}=\frac{\omega }{2a}\left( \tau
-\tau _{b}\right) +\frac{\kappa }{b}.  \label{n12}
\end{equation}
The parameter $m$ is equal to $x_{b}^{2}\,/\,b^{2}\leq 1$ and determines the
constant $\kappa $ on the right-hand side via the complete elliptic integral
of the first kind
\begin{equation}
\kappa =K\left( m\right) =\int\limits_{0}^{1}\frac{dt}{\sqrt{\left(
1-t^{2}\right) \left( 1-m\,t^{2}\right) }}.  \label{n13}
\end{equation}
This constant fixes the period $T$ via formula (\ref{n11}) for $\tau =\tau
_{a}$ as follows
\begin{equation}
2\kappa =\frac{\omega }{2a}bT.  \label{n14}
\end{equation}
The general solution of Eq.~(\ref{n12}) is
\begin{equation}
x_{\text{{\rm cl}}}\left( \tau ,\tau _{b},m\right) =x_{b}\,{\rm sn}\left(
z\left( \tau \right) ;m\right) ,  \label{n15}
\end{equation}
where $z\left( \tau \right) =\omega b\left( \tau -\tau _{b}\right)
\,/\,2a+\kappa ,$ so that $z_{b}=\kappa ,\,\,z_{a}=-\kappa $ and ${\rm sn}%
\left( z;m\right) $ is the elliptic function running from $-1$ to $1$ for $%
\tau \in (\tau _{a},\tau _{b})$, thus ensuring the correct boundary
conditions $x_{\text{{\rm cl}}}\left( \tau _{b}\right) =x_{b},\,x_{\text{%
{\rm cl}}}\left( \tau _{a}\right) =-x_{b}.$

According to Eqs.~(\ref{Kg}) and (\ref{Ome@}), the fluctuations $\delta
x\left( \tau \right) =y\left( \tau \right) $ around the solution (\ref{n15})
are governed by the differential operator $K_{1}\left( \tau \right) =$ $%
-d^{2}\,/\,d\tau ^{2}+\omega ^{2}\left( 3x_{\text{{\rm cl}}%
}^{2}-a^{2}\right) \,/\,2a^{2}$. The boundary conditions are eventually
irrelevant for the level splitting in the ground state, since this will
require taking the limit of an infinite time interval. As an example, we
consider here Dirichlet boundary conditions for eigenfunctions of the
operator $K_{1}\left( \tau \right) :y_{b}=y_{a}=0.\,$ The derivation of
fluctuation determinant requires, in general, two independent solutions of
the homogeneous differential equation which after going over from the time $%
\tau $ to the variable $z(\tau )$ takes the Lame's form
\begin{equation}
\stackrel{\cdot \cdot }{h}\left( z\right) +\left[ \frac{2a^{2}}{b^{2}}-6m\,%
{\rm sn}^{2}\left( z;m\right) \right] h\left( z\right) =0.  \label{n16}
\end{equation}
By translational invariance, the first independent solution $\eta \left(
t\right) $ to this equation is the derivative $\left( \partial /\partial
\tau _{b}\right) x_{\text{{\rm cl}}}(z(\tau );m)$ of Eq.~(\ref{n15}).
Normalizing, we have explicitly
\begin{eqnarray}
\eta \left( \tau \right) &=&N\frac{\partial x_{\text{{\rm cl}}}\left( \tau
,\tau _{b},m\right) }{\partial \tau _{b}}  \nonumber \\
&=&-N\frac{\omega }{2a}b\,x_{b}\,{\rm cn}\left( z;m\right) \,{\rm dn}\left(
z;m\right) ,  \label{n17}
\end{eqnarray}
whose time derivative is
\begin{eqnarray}
\stackrel{\cdot }{\eta }\left( \tau \right) &=&N\left( \frac{\omega }{2a}%
b\right) ^{2}\,x_{b}\,{\rm sn}\left( z;m\right)  \nonumber \\
&&\times \left[ {\rm dn}^{2}\left( z;m\right) +m\,{\rm cn}^{2}\left(
z;m\right) \right] .  \label{n18}
\end{eqnarray}
Here ${\rm cn}\left( z;m\right) $ and ${\rm dn}\left( z;m\right) $ are the
elliptic functions. The normalization factor $N$ is determined by the
condition (\ref{norm@}) as follows
\begin{equation}
N^{-2}=\,x_{b}^{2}\,\left( \frac{\omega }{2a}b\right)
\,\int\limits_{-k}^{k}dz\,{\rm cn}^{2}z\,{\rm dn}^{2}z\,.  \label{n19}
\end{equation}
Performing the integral yields
\begin{eqnarray}
&&\!\!\!\!\!\!\!\!\!\!\!\!\!\!\!\!\!N^{-2}=\,-\frac{4a^{2}}{3\left(
m+1\right) }\,\left( \frac{\omega }{2a}b\right)  \nonumber \\
\times &&\left[ \left( 1-m\right) \kappa -\left( m+1\right) \varepsilon
\right] \,,  \label{n20}
\end{eqnarray}
where $\varepsilon =E(m)$ is given by the complete elliptic integral of the
second kind
\begin{equation}
E\left( m\right) =\int\limits_{0}^{1}dt\sqrt{\frac{1-mt^{2}}{1-t^{2}}}.
\label{n20bis}
\end{equation}

The solution (\ref{n17}) is a zero mode of the operator $K_{1}\left( \tau
\right) ,$ since it satisfies Dirichlet boundary condition
\begin{equation}
\eta _{b}=\eta _{a}=-N\frac{\omega }{2a}b\,x_{b}\,{\rm cn}\,\kappa \,{\rm dn}%
\,\kappa =0.  \label{n21}
\end{equation}
Because of the property (\ref{n21}), the Dirichlet determinant (\ref{GY@})
vanishes, so that we may only calculate the primed determinant according to
formula (\ref{5.13}). This does not require the second independent solution $%
\xi \left( \tau \right) $ of Eq.(\ref{n16}). From Eq.(\ref{n18}) we observe
\begin{equation}
\stackrel{\cdot }{\eta }_{b}=-\,\stackrel{\cdot }{\eta }_{a}=N\left( \frac{%
\omega }{2a}b\right) ^{2}x_{b}\left( 1-m\right) .  \label{n22}
\end{equation}
Inserting (\ref{n22}) with (\ref{n20}) into Eq.(\ref{5.13}), we find
immediately
\begin{eqnarray}
&&\!\!\!\!\!\!\!\!{\rm Det}^{\prime }K_{1}=-\frac{1}{\dot{\eta}_{b}\dot{\eta}%
_{a}}  \nonumber \\
&=&\frac{4a^{2}}{3x_{b}^{2}}\frac{\left[ (m+1)\,\varepsilon -(1-m)\kappa
\right]} {\left( m+1\right) \left( 1-m\right) ^{2}}\bigg /\left(\frac{\omega
}{2a}b\right) ^{3}\,.  \label{n23}
\end{eqnarray}

Let us turn now to the limit of an infinite time interval where $%
E\rightarrow 0$ and $x_{b}$ and $b$ go to the constant $a,$ the parameter $m$
tends to unity as $\left( 1-m\right) \rightarrow 16\,\exp \left( -2\kappa
\right) $, and $\varepsilon \rightarrow 1.$ Using Eq.~(\ref{n14}), we obtain
from (\ref{n23})
\begin{equation}
{\rm Det}^{\prime }K_{1}\rightarrow \frac{e^{4\kappa }}{24\omega ^{3}}=\frac{%
e^{\omega T}}{24\omega ^{3}}  \label{n24}
\end{equation}
In the same limit
\begin{equation}
{\rm sn\,}\left[ \frac{\omega }{2a}b\left( \tau -\tau _{b}\right)
+k;\,1\right] \rightarrow \tanh \left[ \frac{\omega \left( \tau -\tau
_{0}\right) }{2}\right] \,,  \label{n25}
\end{equation}
so that (\ref{n15}) reduces to the limiting kink solution (\ref{n9}), for
which the fluctuation determinant is (\ref{n24}). The presence of
exponentially divergent factor $e^{\omega T}$ like in Eq.(\ref{n24}) is a
special future of fluctuation determinants in the limit $T\rightarrow \infty
.$ It also appears if one derives the determinant of harmonic differential
operator $\tilde{K}=K_{1}^{\omega }=-d^{2}/d\tau ^{2}+\omega ^{2},$ which
governs the fluctuations around the trivial constant classical solution $x_{%
{\rm cl}}\left( \tau \right) =a,$ with the same Dirichlet boundary
conditions. Indeed, inserting of the independent solutions $\eta \left( \tau
\right) =\cosh \omega \tau ,\,\xi \left( \tau \right) =\sinh \omega \tau $
into Eq.~(\ref{GY@}) yields directly
\begin{equation}
{\rm Det}\,K_{1}^{\omega }=\frac{\sinh \omega T}{\omega }\rightarrow \frac{%
e^{\omega T}}{2\omega },  \label{n26}
\end{equation}
where the right-hand side being the large-$T$ limit. Certainly, when
considering the more relevant ratio of (\ref{n24}) and (\ref{n26}), we
obtain the finite result
\begin{equation}
{\rm Det}^{\prime }K_{1}\Big /{\rm Det}\,K_{1}^{\omega }\rightarrow \frac{1}{%
12\omega ^{2}}\,,  \label{n27}
\end{equation}
which agrees, of course, with previous calculation \cite{1}.

The primed determinant (\ref{n24}) can also be derived from (\ref{5.13})
using only the asymptotic behavior of the independent solutions $\eta \left(
\tau \right) $ and $\xi \left( \tau \right) $ at $T\rightarrow \infty $ \cite
{15,1}. For this purpose, we set the particle energy in (\ref{n4}) equal to
zero. The the elliptic functions degenerate into hyperbolic, simplifying Eq.(%
\ref{n16}) to
\begin{equation}
\stackrel{\cdot \cdot }{h}\left( z\right) -2\left( 2-3\cosh ^{-2}z\right)
h\left( z\right) =0,  \label{n28}
\end{equation}
\newline
where $z\left( \tau \right) =\omega \left( \tau -\tau _{0}\right) /2$ with $%
\tau _{0}=\left( \tau _{b}+\tau _{a}\right) /2.$ Now we are looking for the
asymptotics of two independent solutions to this equation at $\tau _{b}-\tau
_{a}\rightarrow \infty $. For the solution corresponding to (\ref{n17}) we
find
\begin{eqnarray}
&&\eta \left( \tau \right) =-N\;\frac{a\omega }{2}\;\cosh ^{-2}z %
\mathop{\longrightarrow }_{\tau _{b}-\tau _{a}\rightarrow \infty } -2a\omega
N\;e^{-2\left| z\right| },  \label{n29}
\end{eqnarray}
the proper normalization factor being $N^{-2}=2a^{2}\omega /3$. For the
second independent solution, the asymptotic behavior may be deduced from the
constancy of Wronskian $W(\eta ,\xi )$ as follows
\begin{equation}
\xi \left( \tau \right) \mathop{\longrightarrow }_{\tau _{b}-\tau
_{a}\rightarrow \infty } \pm e^{2\left| z\right| }.  \label{n30}
\end{equation}
Here the normalization is irrelevant since the expressions (\ref{5.13}) is
independent of it. The solutions (\ref{n29}) and (\ref{n30}) have the
Wronskian $W(\eta ,\xi )=-4a\omega ^{2}N$, and the asymptotic boundary
conditions for $\eta \left( \tau \right) $%
\begin{eqnarray}
&&~\eta _{b}=\eta _{a}\approx -2a\omega N\;e^{-\omega (\tau _{b}-\tau
_{a})/2}, ~~ \dot{\eta }_{b}=-\dot{\eta }_{a} \approx\omega \eta _b,
\label{n31}
\end{eqnarray}
and for $\xi \left( \tau \right) $%
\begin{equation}
\xi _{b}=-\xi _{a}\approx e^{\omega (\tau _{b}-\tau _{a})/2},\quad \stackrel{%
\cdot }{\xi }_{b}=\;\dot{\xi }_{a}\approx \omega \xi_b.  \label{n32}
\end{equation}
Inserting this into formula (\ref{5.13}), with the right-hand side rewritten
as $-\xi_a\xi_b/W^2$, we obtain once again the result (\ref{n24}).

{}~\newline
Acknowledgment:\newline
We thank Drs.~A.~Pelster and S.~Shabanov for useful discussions.

\end{document}